\documentclass[twocol]{ametsocV6.1}

\usepackage{amsmath}
\usepackage{graphicx} 
\usepackage{subcaption}
\usepackage{subfloat}

\title{Estimating Vertical Velocity in Convective Updrafts from Temperature, Pressure, and Latent Heating}

\authors{Amel Derras-Chouk,\aff{a,b}\correspondingauthor{Amel Derras-Chouk, ad4429@columbia.edu}
Gregory Elsaesser,\aff{b,c}
Zhengzhao Johnny Luo,\aff{d}
Toshi Matsui,\aff{e,f}
Andreas F. Prein\aff{g}, and
Jingbo Wu\aff{b,h}
}

\affiliation{\aff{a}{Columbia Climate School, Columbia University, New York, NY, USA}\\
\aff{b}{NASA Goddard Institute for Space Studies, New York, NY, USA}\\
\aff{c}{Department of Applied Physics and Mathematics, Columbia University, New York, NY, USA}\\
\aff{d}{Department of Earth and Atmospheric Sciences, City College of New York, CUNY, New York, New York, USA}\\
\aff{e}{Mesoscale Atmospheric Processes Laboratory, NASA Goddard Space Flight Center, Greenbelt, MD, USA}\\
\aff{f}{Earth System Science Interdisciplinary Center – ESSIC, University of Maryland, College Park, MD, USA}\\
\aff{g}{Institute for Atmospheric and Climate Science, ETH Zürich, Zurich, Switzerland}\\
\aff{h}{Center for Climate Systems Research (CCSR), The Earth Institute, Columbia University, New York, New York, USA}
}

\abstract{The vertical velocity in convective clouds ($w_c$) mediates convective anvil development and global moisture transport, influencing Earth's energy budget, but has yet to be estimated globally over long periods due to the absence of spaceborne retrievals. Here, a method for estimating $w_c$ given vertical profiles of in-cloud temperature, pressure, and latent heating rate is presented and assessed. The method relies on analytical models for the approximately linear relationship between $w_c$ and condensation rate ($\dot{q}_{vc}$) in convective clouds, which we derive from steady-state and non-steady-state plume models. We include in our analysis a version of $\dot{q}_{vc}/w_c$ derived from the supersaturation rate in convective clouds, recently presented in \citet{kukulies_simulating_2024}. We assess the accuracy of $w_c$ estimates against convective cloud simulations run with different model cores and spatial resolutions in both tropical and mid-latitude environments. The velocity estimates exhibit lower uncertainties and higher precision in the tropics than they do in the mid-latitudes. Vertical velocity is estimated to within $\approx1$ m/s for most samples in the tropics. Potential applications, validation against future satellite mission retrievals, and approaches for improving the estimation are discussed.}

\begin{document}

\maketitle

%
%
%
\statement
	A method for estimating the vertical velocity ($w_c$) in clouds using readily available satellite data products is proposed and evaluated, providing a way to quantify for the first time $w_c$ over broad temporal and spatial scales. The results of this study demonstrate that the method's vertical velocity estimates are within a few m/s of the true values with relatively high precision. By estimating $w_c$ with available satellite data, longer-term historical records of convective vertical velocity and large-scale dynamics can be generated. 

%


\section{Introduction}

Convective clouds develop through complex interactions between thermodynamic and dynamic processes on a wide range of spatial and time scales, making them difficult to model and fully observe. For this reason, simplifications are often made when modeling convective clouds. In global climate models, for example, cumulus parameterizations account for clouds' influence below the typical 100 km grid spacings used to model climate \citep{arakawa_cumulus_2004, plant_review_2010, rio_ongoing_2019}.

Despite their complexity, convective clouds tend to exhibit a linear relationship between their updraft vertical wind speeds, or vertical velocities ($w_c$), and water vapor phase transition rates ($\dot{q}_{vc}$), which we refer to as condensation rates for the remainder of the text. Detailed analyses of large-eddy simulations (LESs) of shallow liquid-water and mixed-phase deep convective clouds confirm the robustness of this linearity across a range of simulation grid sizes \citep{grant_linear_2022, stephens_distributed_2020, tao_relating_2022, kogan_estimating_2022}. 

Identifying an analytical model for $\dot{q}_{vc}/w_c$ that depends on available satellite data would allow for global and long-term observational estimates of vertical velocity, filling a gap in remote sensing retrievals. The existing satellite retrievals cover a wide wide range of processes, ranging from cloud-top brightness temperatures in, for example, the Geostationary Operational Environmental Satellites (GOES), to optical depth measured by Moderate Resolution Imaging Spectroradiometer (MODIS). The focus of the rest of the paper will be on leveraging latent heating to estimate of in-cloud vertical motion. Latent heating retrievals from the Tropical Rainfall Measuring Mission (TRMM) and the Global Precipitation Measurement Mission (GPM) collectively provide data from 1997 to the present \citep{tao_trmm_2016}. Because of the importance of diabatic heating in atmospheric processes, there have been attempts to further extend this record by using geostationary satellite data to estimate latent heating based on lookup tables \citep{lee_latent_2022}. 

In prior work focusing on the approximate linearity between $w_c$ and $\dot{q}_{vc}$, the authors explained their results using the supersaturation rate of a rising moist parcel \citep{kogan_estimating_2022, grant_linear_2022, heymsfield_cirrus_1975, korolev_effect_2008, korolev_supersaturation_2003}. The supersaturation rate can be expressed as
\begin{align*}
\frac{ds}{dt} = c_1 w_c - c_2 \, C,
\end{align*}
where $s$ is supersaturation, $c_1$ and $c_2$ are temperature- and pressure-dependent functions, and $C$ is condensation rate \citep{squires_growth_1952,lamb_physics_2011}. This expression predicts that $w_c$ and $C$ are linearly related when $ds/dt=0$. In prior work, analysis using this equation often assumes that the vertical velocity is constant \citep{rogers_elementary_1975,pinsky_supersaturation_2013}. In recent work that presents a method to estimate precipitation efficiency from vertical velocity, temperature, and pressure, \citet{kukulies_simulating_2024} derived an expression relating condensation rate to vertical velocity by assuming a supersaturation tendency equal to zero, which reduces to
\begin{align*}
\alpha_{KPM} \equiv -\frac{\dot{q}_{vc}}{w_c} \frac{L_v}{g} = - L_v \frac{ \frac{1}{c_p} \frac{dq_v^{*}}{dT_c} - \frac{q_v^{*} \rho}{p - e_s} }{1 +  \frac{L_v}{c_p}\frac{dq_v^{*}}{dT_c}}. \tag{KPM24} \label{eq:kpm24}
\end{align*}
Here, $g$ (m s$^{-2}$) is the gravitational acceleration, $c_p$ (J kg$^{-1}$ K$^{-1}$) is the specific heat of dry air, $q_v^{*}$ (kg kg$^{-1}$) is the saturated water vapor mixing ratio, $p$ (Pa) is the atmospheric pressure, $e_s$ (Pa) is the saturated water vapor pressure, $L_v$ (J kg$^{-1}$) is the latent heat of vaporization of water, and $T_c$ (K) is the in-cloud temperature. (Note that they derive an expression for the rate of change of mixing ratio, which is approximately equal to the specific humidity. In the subsequent derivations, we consider the rate of change of specific humidity.) The authors found that this model reliably estimated column-integrated condensation rate and precipitation efficiency in simulations of mesoscale convective systems. In another study, \citet{romps_response_2011} examined the response of precipitation to surface warming and suggested that condensation is related to vertical velocity through the one-dimensional advection equation, approximated as
\begin{align*} 
\frac{\dot{q}_{vc}}{w_c} \approx - \frac{dq_{vc}}{dz}. \tag{R11} \label{eq:r11}
\end{align*}
In the subsequent analysis, we include Eq. \eqref{eq:kpm24}, since the model successfully quantified the total phase transition rate of water vapor in simulations of convection. Because Eq. \eqref{eq:r11} is intended as an approximation across a large-scale domain comprising multiple cloud types, we do not use it hereafter, but note it as an alternative way to explain the approximate linearity between $\dot{q}_{vc}$ and $w_c$. 

With this paper, we add to the existing literature new expressions for $\dot{q}_{vc}/w_c$ that diverge from the prior work in that they start from a one-dimensional plume model and permit both steady-state and non-steady-state assumptions. Section \ref{sec:theory} details these derivations. In Section \ref{sec:estimating_wc}, we assess the accuracy to which $w_c$ is estimated in a series of convection simulations using all three analytical expressions. We describe the simulations, the method for identifying convection, and present the results. In Section \ref{sec:discussion} we discuss implications of this work and its limitations, as well as the impact that theoretical assumptions have on the results. We summarize our main conclusions in Section \ref{sec:summary}.

\section{Theory} \label{sec:theory}

We derive expressions to model the proportionality between vertical velocity and condensation rate based on a one-dimensional plume model of a convective cloud. While the model is a major simplification of cloud physics, versions of it have been successfully used to derive simple physical expressions that provide insight about atmospheric behavior. One-dimensional plume models have been used to derive an equation for the relative humidity of the atmosphere \citep{romps_analytical_2014} and to quantify how convective updrafts are expected to change with changing ambient humidity \citep{singh_influence_2013}. Unlike analysis of the supersaturation rate in convective clouds, a plume model has, to our knowledge, never been applied to explain the proportionality between vertical velocity and condensation rate. 

We make several simplifying assumptions in applying a one-dimensional plume model. By reducing the cloud to a one-dimensional system, the model assumes that the cloud has homogeneous properties at each height. This is not strictly true for clouds that occur in nature, where, for example, a process such as entrainment is bound to lead to horizontal inhomogeneity \citep{de_rooy_entrainment_2013}. Moreover, the plume model is derived assuming homogeneous lateral entrainment, despite it having been shown that inhomogeneous entrainment is required to reproduce observed distributions in cloud droplet size spectra \citep{baker_influence_1980, paluch_mixing_1984}. In our formulation of the plume model, we neglect the impact of condensate mass, assuming pseudoadiabatic ascent. We also neglect the pressure perturbation term, assuming that the in-cloud and ambient pressure are equal. 

\subsection{$\dot{q}_{vc}/w_c$ Derived from a Steady-State Plume Model}\label{sec:theory-steady}

The conservation of water vapor specific humidity in an entraining plume is given by
\begin{align}
\frac{\partial (M q_{vc})}{\partial z} &= (q_{va} e - q_{vc} d) - \Phi
\label{eq:cons-qvc}
\end{align}
\citep{houze_jr_cloud_2014}. Here, $M$ is the convective mass flux (kg m$^{-2}$ s$^{-1}$) (sometimes referred to as normalized mass flux),  $q_{vc}$ is the in-cloud specific humidity (kg kg$^{-1}$), $q_{va}$ is the ambient specific humidity (kg kg$^{-1}$), $e$ is the net mass entrainment rate (kg m$^{-3}$ s$^{-1}$), $d$ is the net mass detrainment rate (kg m$^{-3}$ s$^{-1}$), and $\Phi$ captures all microphysical processes that impact the water vapor specific humidity (kg m$^{-3}$ s$^{-1}$). 

Combining this expression with the mass continuity equation, $\partial M/\partial z = e - d$, we find
\begin{align}
M \frac{\partial q_{vc}}{\partial z} = e(q_{va} - q_{vc}) - \Phi.
\end{align}
We define $\dot{q}_{vc} = -\Phi/\rho_c$ (s$^{-1}$), where $\rho_c$ is the in-cloud air density (kg m$^{-3}$), and $\epsilon = e/M$ (m$^{-1}$) to arrive at
\begin{align}
- \frac{\dot{q}_{vc}}{w_c} = \epsilon(q_{va} - q_{vc}) - \frac{\partial q_{vc}}{\partial z}. \label{eq:cons-rh}
\end{align}
This expression relates $\dot{q}_{vc}$ to the in-cloud specific humidity, the ambient specific humidity, and the entrainment rate. 

We derive an expression for the ambient specific humidity starting from the cloud's moist static energy (MSE), $h_c = L_v q_{vc} + g z + c_p T_c$, where $z$ is height (m). While this definition of MSE neglects condensate, the moisture dependence of $c_p$, and the temperature dependence of $L_v$, assumptions that may impact results in some contexts \citep{yano_moist_2017}, we found that the simplest definition was sufficient for our study. Conservation of the cloud's moist static energy requires
\begin{align}
\frac{\partial (M h_c)}{\partial z} = e h_{a} - d h_c.
\end{align}
Here, $h_a=L_v q_{va} + gz + c_p T_a$ is the ambient MSE and $T_a$ is the ambient temperature (K). We insert the expressions for mass continuity, mass flux, and entrainment rate to find
\begin{align}
\frac{\partial h_c}{\partial z} = \epsilon (h_a - h_c). \label{eq:cons_mse_sat}
\end{align}
Taking the derivative of the $h_c$ with respect to height and combining the result with Eq. \eqref{eq:cons_mse_sat} gives
\begin{align}
\epsilon h_a = \epsilon h_c + L_v \frac{\partial q_{vc}}{\partial z} + g + c_p \frac{\partial T_c}{\partial z}.
\end{align}
Substituting the expressions for the ambient and in-cloud moist static energy into the previous equation leads to
\begin{align}
\epsilon(q_{va} - q_{vc}) = & \frac{c_p}{L_v}\epsilon (T_c - T_a) + \nonumber \\
&\bigg[ \frac{\partial q_{vc}}{\partial z} +  \frac{1}{L_v } \left(g + c_p \frac{\partial T_c}{\partial z}\right) \bigg]. \label{eq:qva}
\end{align}

Combining Eqs. \eqref{eq:cons-rh} and \eqref{eq:qva}, we find
\begin{align}
- \frac{\dot{q}_{vc}}{w_c} &= \epsilon \frac{c_p}{L_v} (T_c - T_a) + \frac{1}{L_v} \left(g + c_p \frac{\partial T_c}{\partial z}\right).
\end{align}
We divide by $g/L_v$ everywhere to rewrite this equation in dimensionless form and define
\begin{align}
\alpha_{p}^{steady} \equiv - \frac{\dot{q}_{vc}}{w_c} \frac{L_v}{g} &=  1 + \frac{c_p}{g} \left( \epsilon (T_c - T_a) + \frac{\partial T_c}{\partial z} \right) \label{eq:alpha_steady}.
\end{align}
We refer to this expression as our steady-state model for the ratio of condensation rate to vertical velocity. 

\subsection{Alternative Derivations of Steady-State $\dot{q}_{vc}/w_c$}

We present two alternative ways to derive Eq. \eqref{eq:alpha_steady}, showing that this expression emerges from several unique starting points when common simplifications are applied.

The first method is based on the weak temperature gradient approximation \citep{sobel_weak_2001}, which, as shown in \citet{elsaesser_simple_2022}, can be written 
\begin{align}
\frac{\dot{Q}_{le}}{w_c} = \Gamma_d - \Gamma \label{eq:alpha_elsaesser}
\end{align}
by assuming radiative heating and horizontal advection are negligible. Here, $\dot{Q}_{le}$ is the heating rate due to condensation and eddy diffusion (K/s), $\Gamma = -dT_c/dz$ is the temperature lapse rate in the cloud (K/m), and $\Gamma_d = g/c_{pd}$ is the dry adiabatic lapse rate (K/m). By neglecting the eddy contribution and assuming that the latent heating is due to condensation, we define $\dot{q}_{vc} = - c_{p} \dot{Q}_{le}/L_v$ and find that Eq. \eqref{eq:alpha_elsaesser} is identical to $\alpha_p^{steady}$ with entrainment set to zero.

Alternatively, conservation of energy gives
\begin{align}
\dot{Q} = c_p \frac{dT_c}{dt} - \frac{1}{\rho_c} \frac{d p}{d t},
\end{align}
where $\dot{Q}$ is the total heating rate of a rising parcel (J kg$^{-1}$ s$^{-1}$), and $p$ is pressure (Pa). This version of the conservation of energy is for a steady-state, closed system. We can substitute $dT_c/dt = (dT_c/dz) (dz/dt)$ and quasi-hydrostatic balance for $dp/dt$ to find
\begin{align}
\frac{\dot{Q}}{w_c} \frac{1}{g} = 1 + \frac{c_p}{g} \frac{dT_c}{dz}. \label{eq:cons_energy}
\end{align}
If we assume that the total heating rate is due to condensation, we can replace $\dot{Q}$ by $-\dot{q}_{vc} L_v$. We then find
\begin{align}
-\frac{\dot{q}_{vc}}{w_c} \frac{L_v}{g} = 1 + \frac{c_p}{g} \frac{dT_c}{dz}.
\end{align}
This expression is equivalent to Eq. \eqref{eq:alpha_steady} with entrainment set to zero. 

Eq. \eqref{eq:alpha_elsaesser} shows that vertical velocity is linearly related to latent and eddy diffusion heating, while Eq. \eqref{eq:cons_energy} shows that vertical velocity is linearly related to the total diabatic heating rate. These expressions imply that $w_c$ is linearly related to heating rate, rather than the condensation rate itself. This suggests that $w_c$ may be linearly related to the latent heating rate of the cloud even when ice is present and deposition is the primary heating source, even though our plume-based derivation neglected ice. Eq. \eqref{eq:cons_energy} suggests that the vertical velocity is linearly proportional to the radiative heating rate as well, since this expression is agnostic to the type of heating. 

It is also possible to derive $\alpha_p^{steady}$ using a rising parcel framework detailed in \citet{peters_generalized_2022}. In the supplementary material, we present this approach, which involves fewer assumptions because it accounts for the temperature dependence of the latent heat of vaporization and includes ice, for example. Because this model ultimately performs as well as $\alpha_p^{steady}$ derived above, we omit it from the main text and continue with a simpler analytical formula.

\subsection{$\dot{q}_{vc}/w_c$ Derived from a Non-Steady-State Plume Model}

The conservation of an arbitrary quantity, $\varphi_c$, in a plume is more generally written as
\begin{align}
\frac{D \varphi_c}{D t} = \left( \frac{D\varphi_c}{Dt} \right)_S + \bar{\epsilon} (\varphi_a - \varphi_c),
\end{align}
where $D/Dt = \partial/\partial t + w_c \partial/\partial z$, $\varphi_a$ is the value of the quantity in the non-cloudy air, and $\bar{\epsilon}$ is entrainment. We have neglected horizontal advection terms in the definition of $D/Dt$. (In the derivation shown in Section 2a, the steady state assumption is made by setting $\partial/\partial t$ to zero.) The subscript $S$ refers the value of the quantity when the plume is not entraining. The conservation of in-cloud water vapor becomes
\begin{align}
\frac{D q_{vc}}{D t} = \left(\frac{Dq_{vc}}{Dt} \right)_S + \bar{\epsilon} (q_{va} - q_{vc}). \label{eq:cons_qva}
\end{align}
We can find $(q_{va} - q_{vc})$ from the conservation of MSE,
\begin{align}
\frac{Dh_c}{Dt} = \bar{\epsilon} (h_a - h_c).
\end{align}
Here we have used $(Dh_c/Dt)_S = 0$. Inserting the definition of moist static energy and isolating the term that contains $(q_{va} - q_{vc})$, we find
\begin{align}
L_v \bar{\epsilon} (q_{va} - q_{vc}) =  &L_v \frac{D q_{vc}}{Dt} + c_p \frac{DT_c}{Dt} + gw_c \nonumber \\ 
& - \bar{\epsilon} \left(c_p (T_a-T_c) \right). \label{eq:cons_mse_isolated}
\end{align}
Substituting Eq. \eqref{eq:cons_mse_isolated} into \eqref{eq:cons_qva} gives
\begin{align}
  -L_v \dot{q}_{vc} &= c_p \frac{D T_c}{D t} + gw_c - \bar{\epsilon} c_p (T_a - T_c) \nonumber \\
 &= c_p \left(\frac{\partial T_c}{\partial t} + w_c \frac{\partial T_c}{\partial z} \right) + gw_c + \bar{\epsilon} c_p (T_c - T_a) \label{eq:cons_qva_v2}.
\end{align} 
In the previous line we have defined $(Dq_{vc}/Dt)_S = \dot{q}_{vc}$. 

The task now is to derive an expression for $\partial T_c/\partial t$. If we assume that the cloud is saturated, an assumption supported by observations \citep{politovich_variability_1988, ditas_aerosols-cloud_2012, yang_new_2019}, we can calculate $\partial T_c/\partial t$ from $D/Dt$ of the saturated water vapor mixing ratio, $q_v^{*} = \varepsilon e_s/p$, where $\varepsilon=0.622$. We find
\begin{align}
  \dot{q}_{vc} &= \varepsilon \left( \frac{1}{p} \frac{D e_s}{Dt} - \frac{e_s}{p^2} \frac{D p}{Dt}  \right) \nonumber \\
   &= q_v^{*} \left( \left[\frac{L_v}{R_v T_c} - 1 \right] \frac{1}{T_c} \frac{\partial T_c}{\partial t} + \left[\frac{\varepsilon L_v}{g T_c} \frac{\partial T_c}{\partial z} + 1 \right] \frac{w_c g}{R_d T_c} \right) \nonumber \\
  &= q_v^{*} \left( f_1 \frac{1}{T_c} \frac{\partial T_c}{\partial t} + f_2 \frac{w_c g}{R_d T_c} \right).
\end{align}
Several assumptions are made to arrive at this result. We used an approximation of the Clausius-Clapeyron equation that neglects the temperature dependence of the latent heat, $de_s/dT_c \approx L_v e_s/(R_v T^2)$. We also assumed hydrostatic balance, which was already assumed when we applied the conservation of moist static energy. We use the approximation $dp/dz = - p/(R_d T_\rho) \approx -p/(R_d T_c)$, which assumes the density temperature is equal to the temperature of the cloud. In the final line we defined dimensionless functions $f_1$ and $f_2$ for simplicity. Isolating $\partial T_c/\partial t$, we find
\begin{align}
  \frac{\partial T_c}{\partial t} &= \frac{\dot{q}_{vc}}{q_v^{*}}\frac{T_c}{f_1} - \frac{f_2}{f_1} \frac{w_c g}{R_d}. \label{eq:dTdt}
\end{align}
Substituting Eq. \eqref{eq:dTdt} into Eq. \eqref{eq:cons_qva_v2} gives 
\begin{align}
 - L_v \dot{q}_{vc} = & c_p \frac{\dot{q}_{vc}}{q_v^{*}}\frac{T_c}{f_1} + w_c \left(- \frac{f_2}{f_1} \frac{c_p g}{R_d} + c_p \frac{\partial T_c}{\partial z} + g\right) \nonumber \\
 & + \bar{\epsilon} c_p (T_c - T_a).
\end{align}
Combining terms and dividing by $g$ to make the equation dimensionless, we arrive at
\begin{align}
  -\frac{\dot{q}_{vc}}{w_c} \frac{L_v}{g}  \left(1 + \frac{c_p}{L_v} \frac{1}{q_v^{*}}\frac{T_c}{f_1} \right) = &-\left( \frac{f_2}{f_1} \frac{c_p }{R_d} - \frac{c_p}{g} \frac{\partial T_c}{\partial z} - 1\right) \nonumber \\
  &+ \frac{\bar{\epsilon}}{w_c} \frac{c_p}{g} (T_c - T_a). \label{eq:alpha_nonsteady0}
\end{align}

Defining the last term in the expression as $f_3$, our expression for the ratio of condensation rate to vertical velocity given by the non-steady-state plume model, which we label $\alpha_p$, is
\begin{align}
  -\frac{\dot{q}_{vc}}{w_c} \frac{L_v}{g}  \left(1 + \frac{c_p}{L_v} \frac{1}{q_v^{*}}\frac{T_c}{f_1} \right) = &- \left( \frac{f_2}{f_1} \frac{c_p }{R_d} -  \frac{c_p}{g} \frac{\partial T_c}{\partial z} - 1\right) + f_3. \label{eq:alpha_nonsteady}
\end{align}

\section{Estimating Vertical Velocity in Convection Simulations} \label{sec:estimating_wc}

Having derived analytical expressions relating $w_c$ to $\dot{q}_{vc}$, we assess whether the mean vertical velocity in convective cores can be estimated using these expressions. As mentioned in the introduction, we include Eq. \eqref{eq:kpm24} in our assessment. Intending to apply this method in the future to satellite data, our validation procedure illustrates how noise in existing satellite sounding retrievals of temperature impact results.

We approximate the rate of change of water vapor relative humidity, a quantity that is not retrieved by satellites, by defining $\dot{q}_{vc} = -c_{p} \dot{Q}_l/L_v$, where $\dot{Q}_l$ is the total latent heating rate (K/s). Doing so allows us to reformulate our expressions so that
\begin{equation}
w_c =  - \frac{L_{v}}{g} \frac{\dot{q}_{vc}}{\alpha} = \frac{c_{pd}}{g} \frac{\dot{Q}_l}{\alpha},
\end{equation}
where $\alpha$ is given by Eq. \eqref{eq:alpha_steady}, \eqref{eq:alpha_nonsteady}, or \eqref{eq:kpm24}. By specifying $\dot{q}_{vc}$ in this way, we assume that only condensation and evaporation contribute to the total heating rate in the simulations. This simplification is an extension of our earlier assumption that the cloud contains only liquid water. We moreover neglect contributions from freezing and melting by making this assumption. Since the latent heat of freezing is an order of magnitude smaller than the latent heat of vaporization or sublimation, we maintain leading order accuracy by neglecting melting.

Because any simple entrainment parameterization is likely insufficient to accurately model simulation mixing and would introduce errors to the leading-order terms that contribute to $w_c$, we set $\epsilon$  and $\bar{\epsilon}$ to 0 in the following results. We discuss the effect of entrainment on vertical velocity estimates in Section \ref{sec:discussion}. 

Satellite sounding retrievals characterize the nearest clear-sky temperatures surrounding convective systems, so there is some ambiguity about the true temperature in the cloud. In later analysis where we validate our method to estimate vertical velocity, we also compute the $w_c$ estimated from temperatures shifted by a small amount, testing the sensitivity of the estimated velocities to the potential uncertainties in temperature. Prior comparisons of spaceborne temperature retrievals to atmospheric soundings indicate that the bias in retrieved temperatures when opaque clouds are present is $\approx -2$ K, on average, while the root mean squared error can reach up to 4 K when the most opaque clouds are present \citep{wong_cloud-induced_2015}. Comparisons of Atmospheric Infrared Sounder (AIRS) temperature (mostly clear-sky) retrievals to the Modern-Era Retrospective analysis for Research and Applications, version 2 (MERRA-2) all-sky temperatures similarly show that retrievals can differ from reanalysis data by $\approx$ 2 K \citep{elsaesser_using_2025}. Based on these prior studies, we chose to perturb the in-cloud temperature by $\pm$5 K to evaluate the temperature sensitivity of our vertical velocity estimators. 

\subsection{Data}

\begin{figure*}[h]
    \centering
    \begin{subfigure}{0.85\textwidth}
      \centering
      \includegraphics[width=\linewidth]{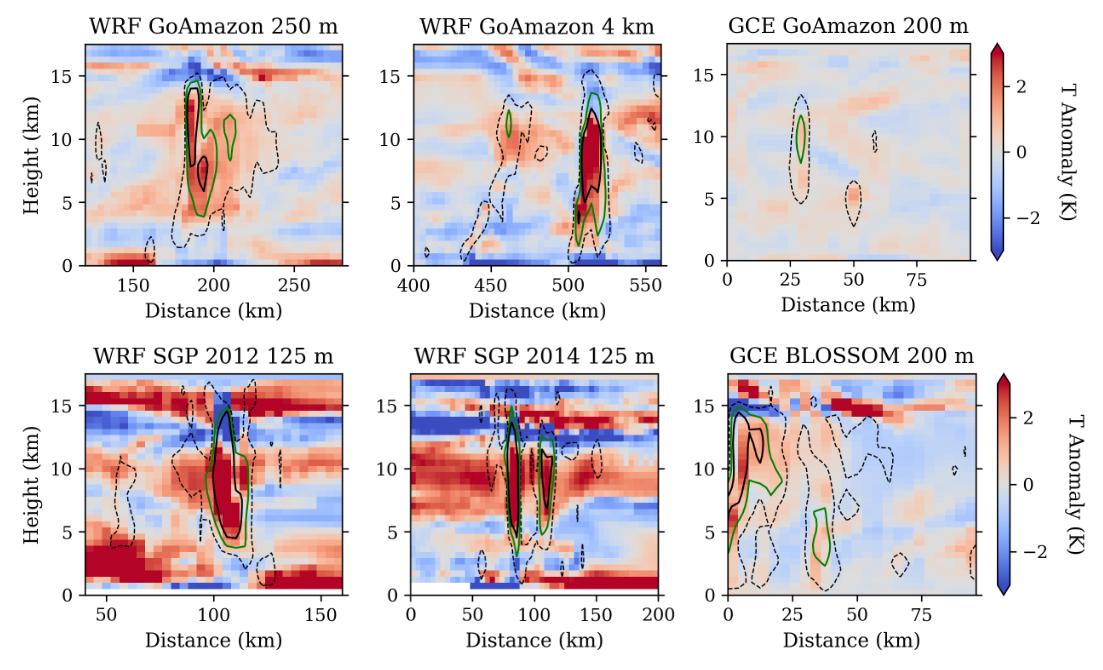}
    \end{subfigure}
    \caption{Temperature anomalies along cross sections of selected domains overlaid with contours of vertical velocity of 1 m s$^{-1}$ (dashed), 5 m s$^{-1}$ (solid green), and 10 m s$^{-1}$ (solid black). All data has been coarsened to 4 km horizontal grid spacing.}
    \label{fig:t_anomalies}
\end{figure*}

\begin{figure*}[h]
    \centering
    \begin{subfigure}{0.85\textwidth}
      \centering
      \includegraphics[width=\linewidth]{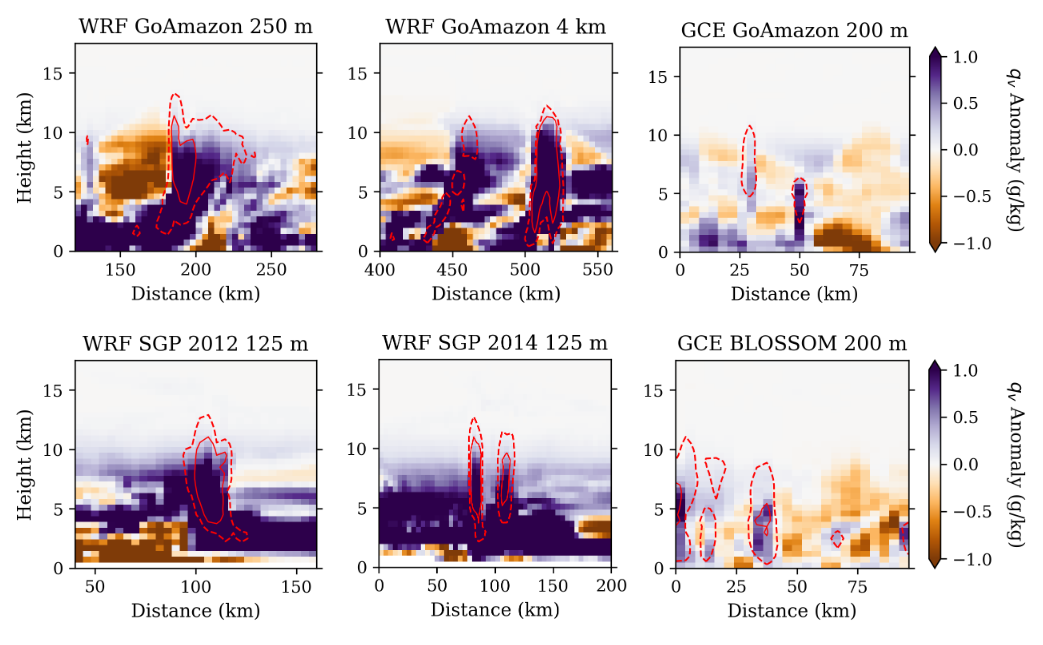}
    \end{subfigure}
    \caption{As in Fig. \ref{fig:t_anomalies}, but for water vapor instead of temperature. The same cross sections are used here. The dashed curves mark latent heating rates of 18 K/hr, while the solid ones mark 90 K/hr.}
    \label{fig:qv_anomalies}
\end{figure*}

We test our vertical velocity estimation method on convective cloud simulations run for a range of grid sizes and generated using two different modeling frameworks. To assess broad applicability of the expressions, we include simulations of both tropical and mid-latitude convection, and both isolated and organized convection.

One set of simulations is based on the Weather Research and Forecasting (WRF) model \citep{skamarock2008time}. The tropical simulations are downscaling ten mesoscale convective systems from the fifth generation ECMWF reanalysis (ERA5) \citep{hersbach2020era5} that were observed during the U.S. DOE ARM Green Ocean Amazon (GoAmazon) field campaign \citep{martin2016introduction}. To test the sensitivity of our $w_c$ estimation method to simulation grid size, which can impact properties like convective aggregation \citep{prein_sensitivity_2021}, we use data that was run with horizontal grid spacings of 250 m, 1 km, 2 km, and 4 km. All of the outputs were conservatively coarsened to a 4 km grid. We refer to the original grid sizes as the ``native" ones in the rest of the paper. While we tested how well $w_c$ is estimated using all four simulations, we include only the 250 m and 4 km cases in the text for brevity. (The results obtained from the 1 and 2 km models are qualitatively similar to the 4 km ones.) Model outputs are saved every 10 minutes, with a domain size of 748 $\times$ 748 km$^2$. The 250 m simulation models 2 hours of an MCS observed on Dec. 10, 2014, while the 4 km simulation spans 36 hours from Dec. 9, 2014 to Dec. 11, 2014. For details of the model setup and parameterizations, see \citet{wang_updraft_2020} and \citet{prein_sensitivity_2021}. We also include two WRF large-eddy simulations (LESs) of mesoscale convective systems observed at the ARM Southern Great Plains (SGP) site on June 15, 2012 and June 12, 2014. These simulations were performed at 125 m grid spacing and coarsened to a 4 km grid, modeling convection over a 752 $\times$ 752 km$^2$ domain. As in the tropical cases, outputs were saved every 10 minutes. All of the WRF simulations use Thompson microphysics.

The second set of simulations are generated with the Goddard Cumulus Ensemble (GCE) model. One of them simulates convection observed during the wet season intensive observation period of the GoAmazon field campaign, detailed in \citet{matsui_thermal-driven_2024}. This data is saved every hour and spans February 14--March 25, 2014. The domain size is 12.8 $\times$ 102.4 km$^2$ and it has 200 m grid spacing. The database contains the collection of isolated tropical deep convection over a few months. The other GCE simulation is of mid-latitude convection observed during the BiLateral Operational Storm-Scale Observation and Modeling (BLOSSOM) project. This was run at 200 m grid spacing and model outputs were saved every 10 minutes over a 24 hour period. The domain size of this simulation is 102.4 $\times$ 102.4 km$^2$. This particular case represents widespread summertime deep convection over the Delmarva Peninsula. All of the GCE simulations used Goddard one-moment 4ICE microphysics \citep{lang_benefits_2014, tao_high-resolution_2016}. Details of this model can be found in \citet{matsui_systematic_2023}. While both GCE simulations are provided with 200 m grid spacings, we conservatively coarsen both of them 4 km to have an analysis grid that is consistent with the WRF simulations.

Despite having comparable spatial resolutions, the two modeling frameworks diverge in their driving methods, physical parameterizations, and dynamic cores. Specifically, the GCE simulations were nudged with homogeneous large-scale forcing and doubly cyclic boundary conditions in an idealized configuration. In contrast, the WRF simulations were driven by initial and lateral boundary conditions within a regional modeling framework. Therefore, evaluating our theory using these different models is essential to determine the robustness of our proposed methods.

Figs. \ref{fig:t_anomalies} and \ref{fig:qv_anomalies} show temperature and water vapor anomalies along cross sections of selected scenes for each simulation. The anomalies are defined as the differences from the domain-mean values at each height. Fig. \ref{fig:t_anomalies} shows shows temperature anomalies as a function of height and contours of vertical velocity, while Fig. \ref{fig:qv_anomalies} shows moisture anomalies as a function of height with contours of latent heating. The contours delineating vertical velocities of 10 m/s surround temperatures that are higher than the rest of the domain and are associated with higher moisture anomalies, although opposite-sign anomalies are noted at higher altitudes (above 12 km) in the mid-latitudes. The temperature anomalies within contours of 1 m/s are less extreme, with negative temperature anomalies even more visible as altitude increases. Large positive latent heating is associated with a more positive water vapor anomaly. This figure illustrates the potential bias that may emerge when the nearest clear-sky temperature (from satellite retrievals) is used to characterize $T_c$. According to the simulations, the temperature differences between the cloud and the environment support our proposed perturbations of 5 K for assessing sensitivities, especially in the regions with $w_c$ greater than 10 m/s. 

\begin{figure*}[h]
\centering
\begin{subfigure}{0.85\textwidth}
  \centering
  \includegraphics[width=\linewidth]{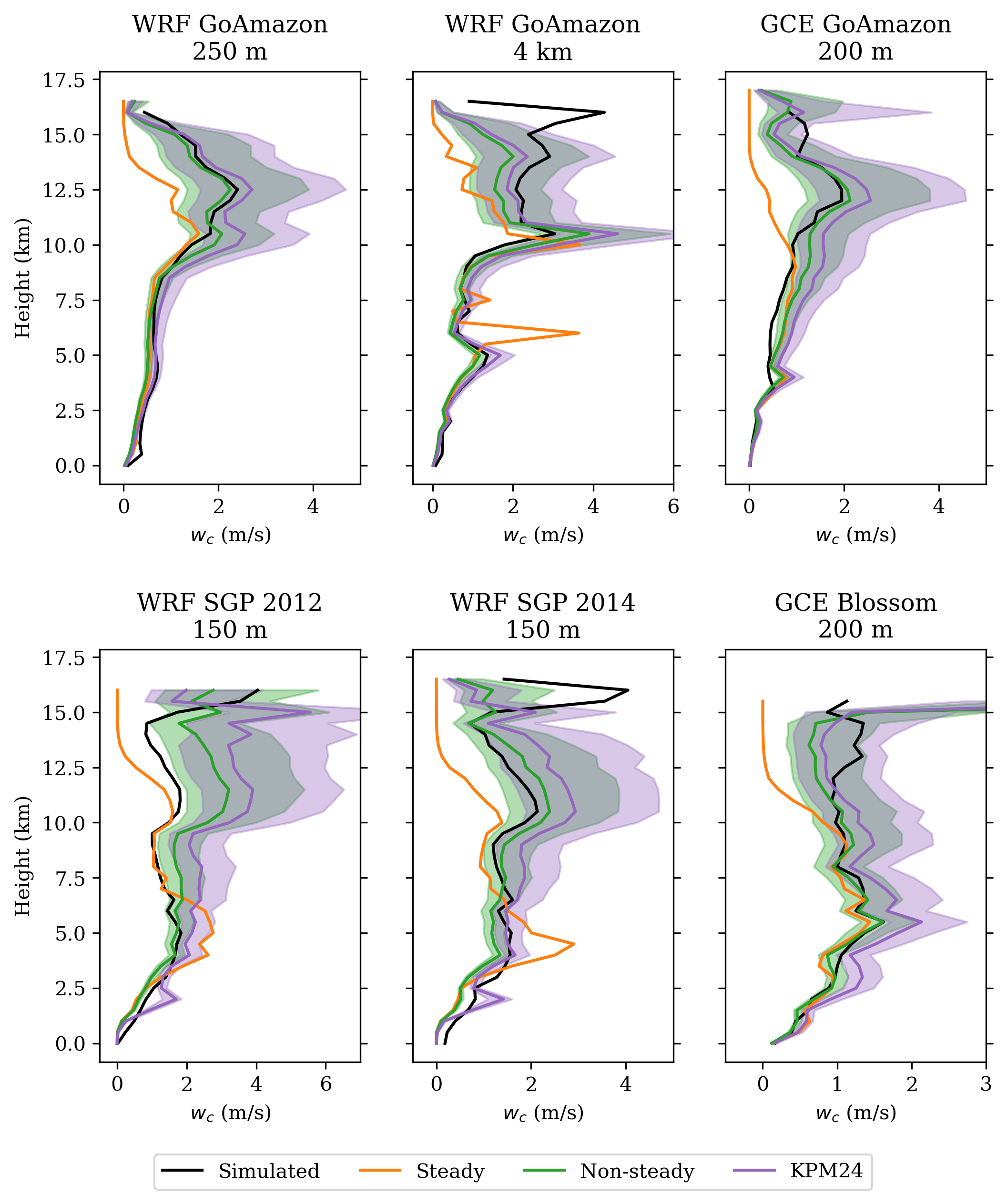}
\end{subfigure}
\caption{Vertical profiles of the mean vertical velocity in convective cores estimated using the $\alpha_p^{steady}$, $\alpha_p$, and $\alpha_{KPM}$ models and the mean simulated velocity profiles. The upper end of the shaded regions marks velocities predicted with an in-cloud temperature 5 K less than the mean core temperature, while the lower end of the shaded regions marks velocities predicted with $T_c$ 5 K higher than the in-core temperature.}
\label{fig:mean_wc_profiles}
\end{figure*}

\subsection{Methods}

\begin{figure*}[h!]
    \includegraphics[width=\linewidth]{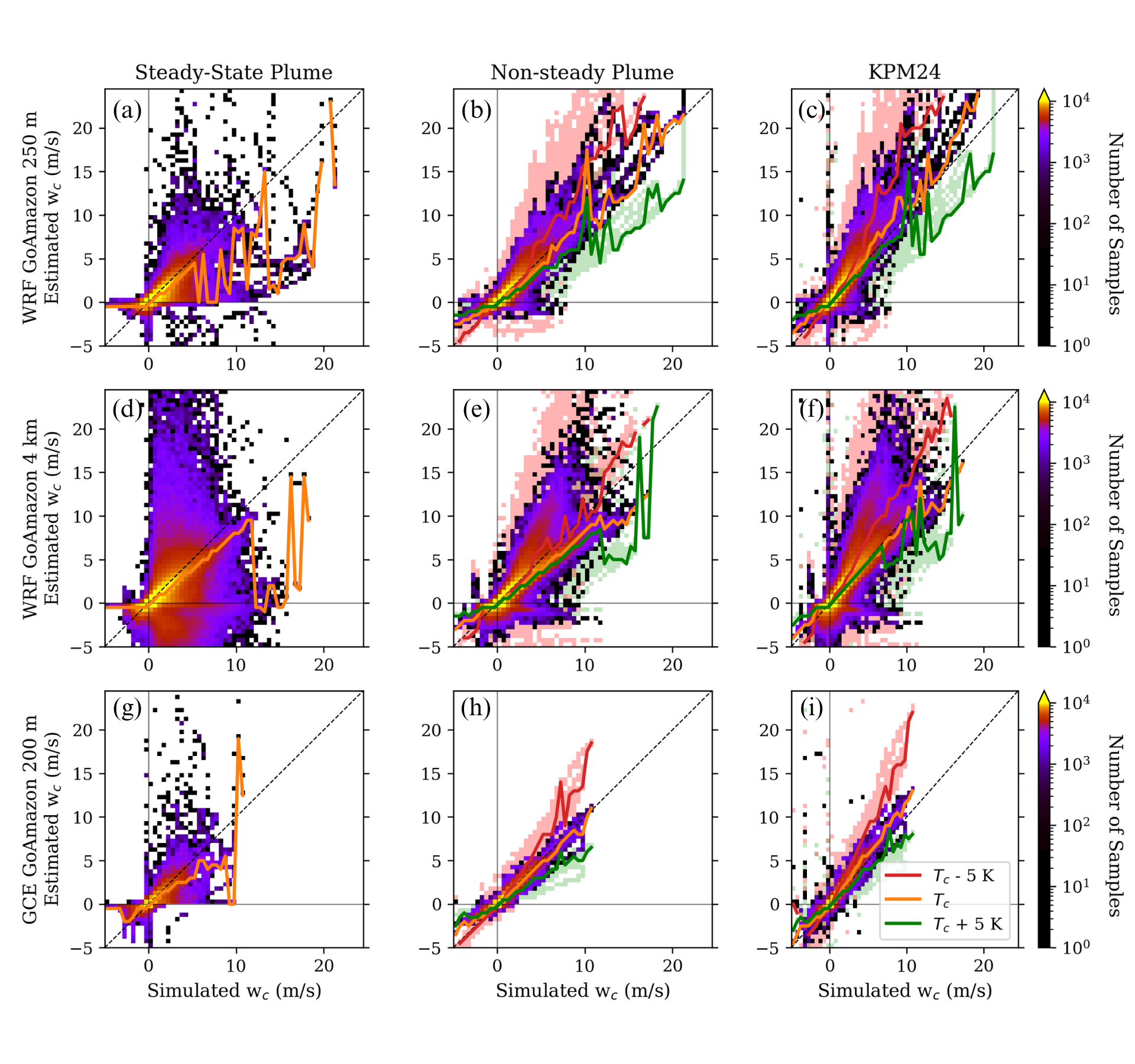}
    \caption{Two-dimensional histograms of estimated versus true vertical velocities given by each version of $\alpha$ for the tropical simulations. In each figure, the orange line delineates the most probable vertical velocity estimate for each true vertical velocity value. The green and red lines show the most probable estimated vertical velocities when the in-cloud temperature increases or decreases by 5 K, respectively. The red and green backgrounds show two-dimensional histograms of estimated vs. true vertical velocities for in-cloud temperatures shifted by 5 K. The velocity bin size is 0.5 m/s.}
    \label{fig:pred_vs_true_a}
\end{figure*}

\begin{figure*}[h!]
  \includegraphics[width=\linewidth]{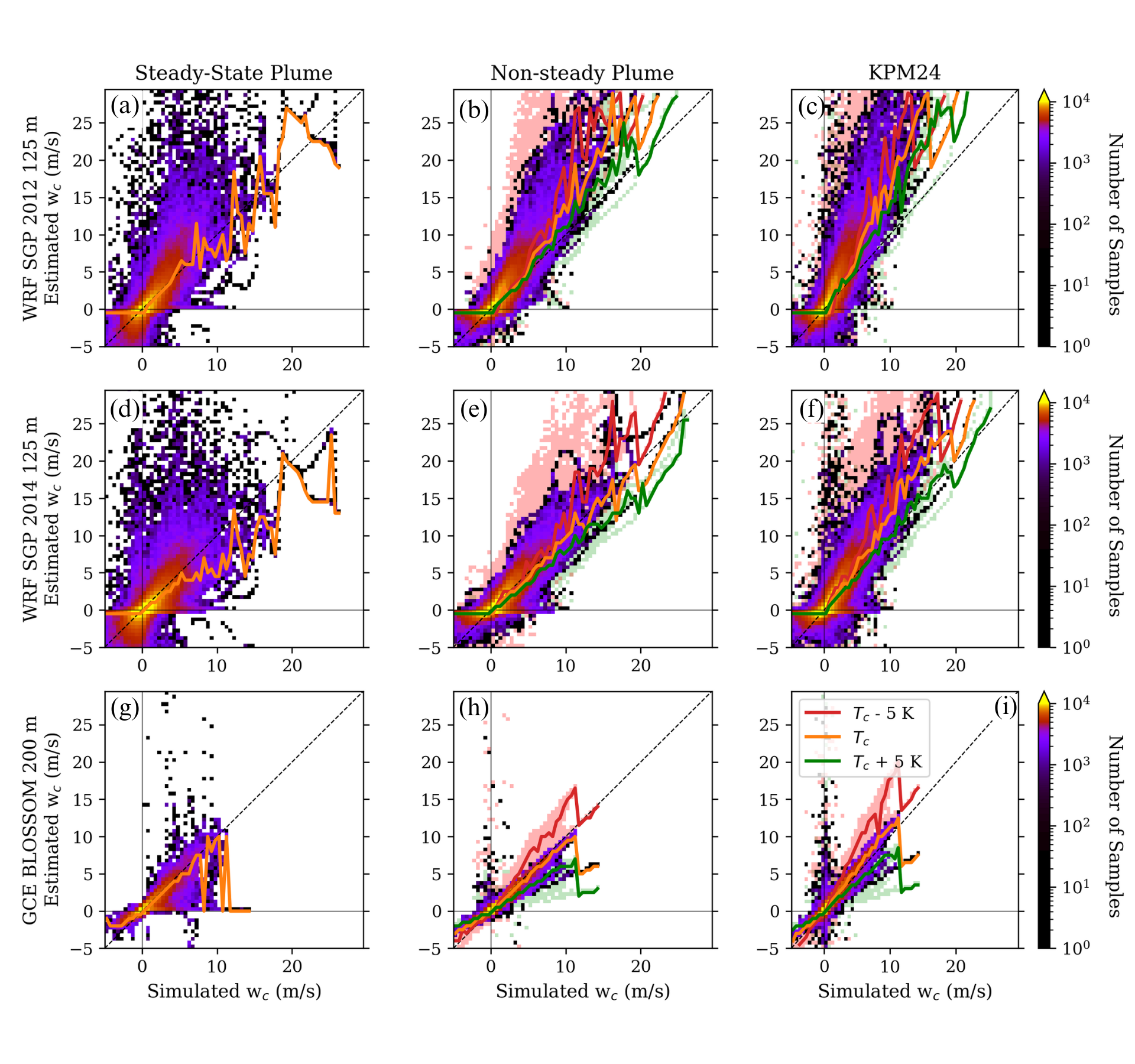}
      \caption{As in Fig. \ref{fig:pred_vs_true_a}, but for the three mid-latitude simulations.}
    \label{fig:pred_vs_true_b}
\end{figure*}

We test whether the average velocities within simulated convective cores are accurately estimated using each version of $\alpha$. In all of the subsequent results, we compute the vertical velocity using the coarsened (4 km) simulation data. 

To identify convective cores in the simulations, we filter for connected pixels with radar reflectivity greater than 10 dBz that are continuous between 2 and 6 km. We require each set of connected pixels to have a 10 dBz echo-top height greater than or equal to 10 km. This method has been applied to CloudSat data to filter for convective cores \citep{jeyaratnam_satellitebased_2021, takahashi_revisiting_2023}. 

We compute the horizontal mean of vertical velocity, temperature, pressure, and latent heating for each convective core, obtaining a vertical profile for each variable. Because the GCE data provides the microphysical phase transition rates, rather than the total latent heating, we compute the total latent heating contributed from all processes, and use that value in the subsequent analysis. For all of the simulations, we compute $dT_c/dz$ for each core from the mean temperature profile. Unless otherwise stated, we remove from our analysis the small portion of updrafts that have negative latent heating, which rarely occur. In doing this, we assume that positive latent heating is associated with upward motion. The mean profile of $w_c$ for each core constitutes our ground truth, which we refer to as the true or simulated $w_c$. 

\subsection{Results}

\subsubsection{Mean Profiles of Vertical Velocity}

Fig. \ref{fig:mean_wc_profiles} shows the mean estimated $w_c$ profiles compared to the true $w_c$ profiles for each simulation. The means are taken over the entire time period and domain spanned by each simulation. For vertical velocities estimated with the non-steady and KPM24 models, the shaded region denotes the $w_c$ envelope computed by perturbing the in-cloud temperature by $\pm5$ K. (Because the steady-state model depends on vertical derivatives of temperature and not $T_c$ itself, we do not need to perform this perturbation.)

For the ranges of domain sizes, native simulation grid sizes, and across regimes, all expressions for $\alpha$ yield average convective core $w_c$ profiles reasonably close to the true profiles. For most altitudes, the $\alpha_{KPM}$ model tends to estimate a higher vertical velocity than $\alpha_p$, while $\alpha_p$ tends to estimate a higher $w_c$ than $\alpha_p^{steady}$. Across all simulations, the temperature sensitivity of vertical velocities estimated with $\alpha_{p}$ and $\alpha_{KPM}$ increases with altitude. The mean vertical velocities estimated with the KPM24 model are the most sensitive to changes in input temperature. Considering that this model was derived from the supersaturation rate of water vapor, which depends on the temperature-sensitive Clausius-Clapeyron equation, this is expected. 

The trends in the mean profiles shown in Fig. \ref{fig:mean_wc_profiles} provide some information about the strengths and weaknesses of each $w_c$ estimator. Velocities estimated with $\alpha_p^{steady}$ are more accurate below $\approx 10$ km across all simulations. Above 10 km, the model tends to underestimate $w_c$. Moreover, the mean profiles of vertical velocity are noisier than the profiles given by the other two estimators, indicating the uncertainty in these $w_c$ estimates may be high. The mean profiles of $w_c$ estimated with the non-steady plume model are broadly more accurate in the tropical simulations than they are in the mid-latitude simulations, although the estimated velocities in the GCE BLOSSOM case maintain high accuracy. In the upper troposphere, the model accurately estimates velocities in the WRF GoAmazon 250 m case, while it underestimates velocities in the WRF GoAmazon 4 km case, providing some evidence of grid-size dependence in the $w_c$ estimation accuracies. The mean $w_c$ profiles estimated with $\alpha_{KPM}$ exhibit trends similar to the ones noted for the $\alpha_p$ estimates, and generally predicts slightly higher velocities than those given by the non-steady plume model.

\subsubsection{Distributions of Vertical Velocity Estimates}

We plot two-dimensional histograms of the estimated versus simulated vertical velocities in Figs. \ref{fig:pred_vs_true_a} and \ref{fig:pred_vs_true_b} for the tropical and mid-latitude simulations, respectively. We have kept all data points in these figures, deciding not to apply the $\dot{Q}_l>0$ filter to the data. 

The vertical velocities estimated with $\alpha_p^{steady}$, shown in the first column of Figs. \ref{fig:pred_vs_true_a} and \ref{fig:pred_vs_true_b}, exhibit a broad spread in estimated $w_c$, even though the most likely vertical velocity is fairly accurate for most of the models. The spread in the estimated vertical velocities explains the noisiness in the mean $w_c$ profiles shown in Fig. \ref{fig:mean_wc_profiles}. These trends are apparent across all simulations. Overall, the results indicate that there is a small negative bias in the estimated vertical velocity obtained from the steady-state model. 

The estimates of $w_c$ given by $\alpha_p$ exhibit a much narrower spread than the ones given by $\alpha_p^{steady}$. Variability among the different simulations leads to different conclusions about the accuracy of the non-steady plume model in estimating vertical velocity. In most of the simulations, the estimated vertical velocity is approximately equal to the true one. In the WRF GoAmazon 4 km case, $\alpha_p$ slightly underestimates $w_c$, while in the WRF SGP 2012 case, it slightly overestimates it. Across all of the simulations, as the true $w_c$ increases, the spread in the range of $w_c$ estimated by shifting $T_c$ by $\pm 5$ K increases. These trends motivate further analysis of the accuracy in vertical velocity estimates with temperature. 

\begin{figure*}[h!]
\centering
\begin{subfigure}{\textwidth}
  \centering
  \includegraphics[width=\linewidth]{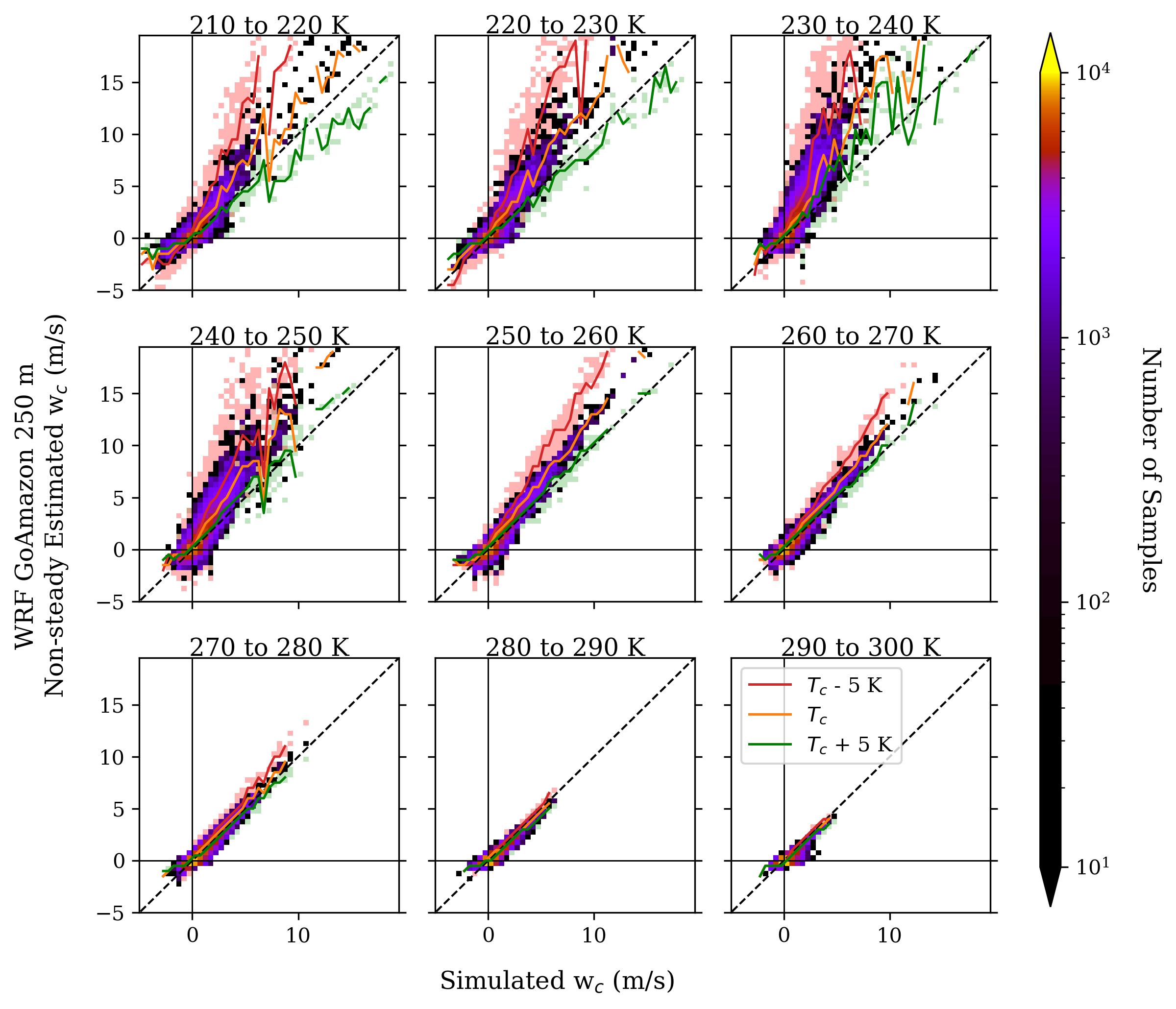}
\end{subfigure}
\caption{Each panel shows a two-dimensional histogram of the vertical velocity estimated with $\alpha_p$ as a function of the simulated $w_c$ for the WRF GoAmazon 250 m simulation subset to different in-cloud temperature bins. The figures are analagous to the ones shown in Fig. \ref{fig:pred_vs_true_a}.}
\label{fig:pred_vs_true_tc_binned_250m}
\end{figure*}

\begin{figure*}[h!]
\centering
\begin{subfigure}{\textwidth}
  \centering
  \includegraphics[width=\linewidth]{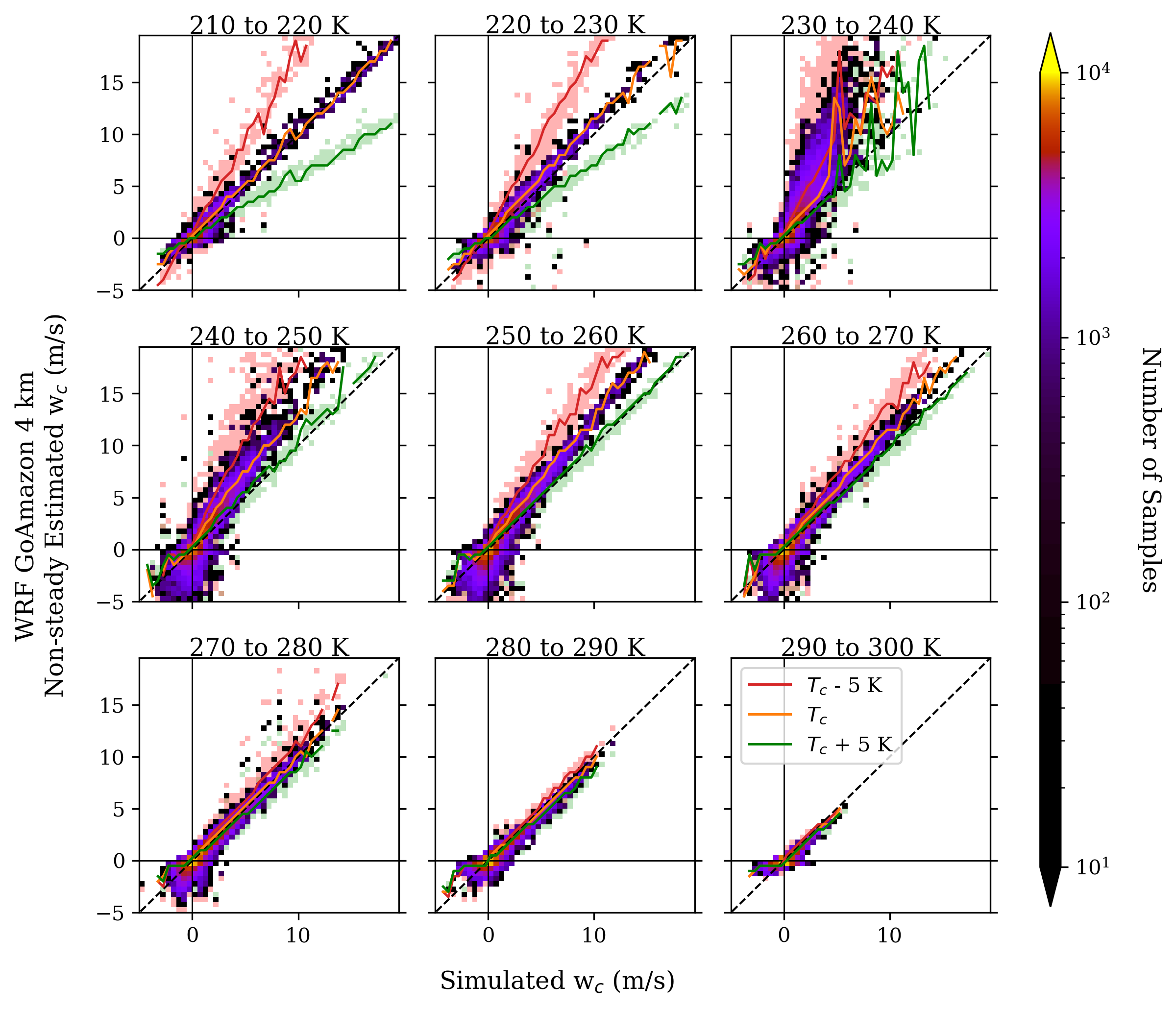}
\end{subfigure}
\caption{As in Fig. \ref{fig:pred_vs_true_tc_binned_250m}, but for the WRF GoAmazon 4 km simulation.}
\label{fig:pred_vs_true_tc_binned_4km}
\end{figure*}

The vertical velocities estimated using $\alpha_{KPM}$ are similarly more precise than the ones estimated with the steady-state plume derivation. As in the $\alpha_p$ case, the variability of the $w_c$ estimates with the different simulations complicates the assessment of this models' skill in estimating $w_c$. Across most of the simulations, the estimated vertical velocity is fairly accurate. In the GCE and SGP, this model tends to overestimate velocity, especially when the true velocities become large. 

For $\lesssim$10\% of data points, the estimated vertical velocities have the opposite sign of the true vertical velocities. This can happen when there is vertical motion in the simulation even though the latent heating rate is negative, which may be related to overshooting tops or negatively buoyant upward-moving air that can sometimes occur in convection \citep{xu_updraft_2001, luo2010}. Conversely, sometimes there is downward motion in the simulation even though the latent heating rate is positive, behavior that has been noted in LESs previously \citep{wang_characteristics_2014}. Our assumption that positive latent heating is accompanied by upward moving air is not strictly true, but it holds in most cases. 

\subsubsection{Impact of Temperature and Model Grid Spacing}

Because we derived expressions relating latent heating to $w_c$ that neglected ice, an unreasonable assumption at high altitudes, we expect that the accuracy of the vertical velocity estimates may depend on whether ice is present. We focus on the estimates given by $\alpha_p$, since, based on Figs. \ref{fig:pred_vs_true_a} and \ref{fig:pred_vs_true_b}, they have higher precision and less bias than the other two estimators. Additionally, to examine the impact of model grid size on the $w_c$ estimators, we focus on WRF GoAmazon 250 m and 4 km ensemble simulations of the same mesoscale convective systems. 

Figs. \ref{fig:pred_vs_true_tc_binned_250m} and \ref{fig:pred_vs_true_tc_binned_4km} show two-dimensional histograms of the estimated versus simulated $w_c$ grouped into 10 K temperature bins for the WRF GoAmazon 250 m and 4 km runs, respectively. Both figures show that $w_c$ estimates are fairly precise for all temperatures except within 230--250 K, where the distributions of estimated vertical velocities broaden. As we expect, the $w_c$ estimates are more accurate when the temperature of the cloud is higher, where our liquid-only assumption is acceptable. Below 230 K, $w_c$ estimates made with $\alpha_p$ are still fairly accurate, albeit noisy in the 250 m case. Moreover, Figs. \ref{fig:pred_vs_true_tc_binned_250m} and \ref{fig:pred_vs_true_tc_binned_4km} illustrate that the temperature sensitivity of the estimated $w_c$, shown in the red and green curves, increases as the in-cloud temperature bin decreases. The range of $w_c$ values expected by shifting $T_c$ by $\pm$5 K is broader in the upper troposphere, leading to high uncertainties in the vertical velocity. This uncertainty is amplified when the true $w_c$ is high.

\subsubsection{Errors in $w_c$ estimates}

\begin{figure*}[h!]
    \centering
     \includegraphics[width=\linewidth]{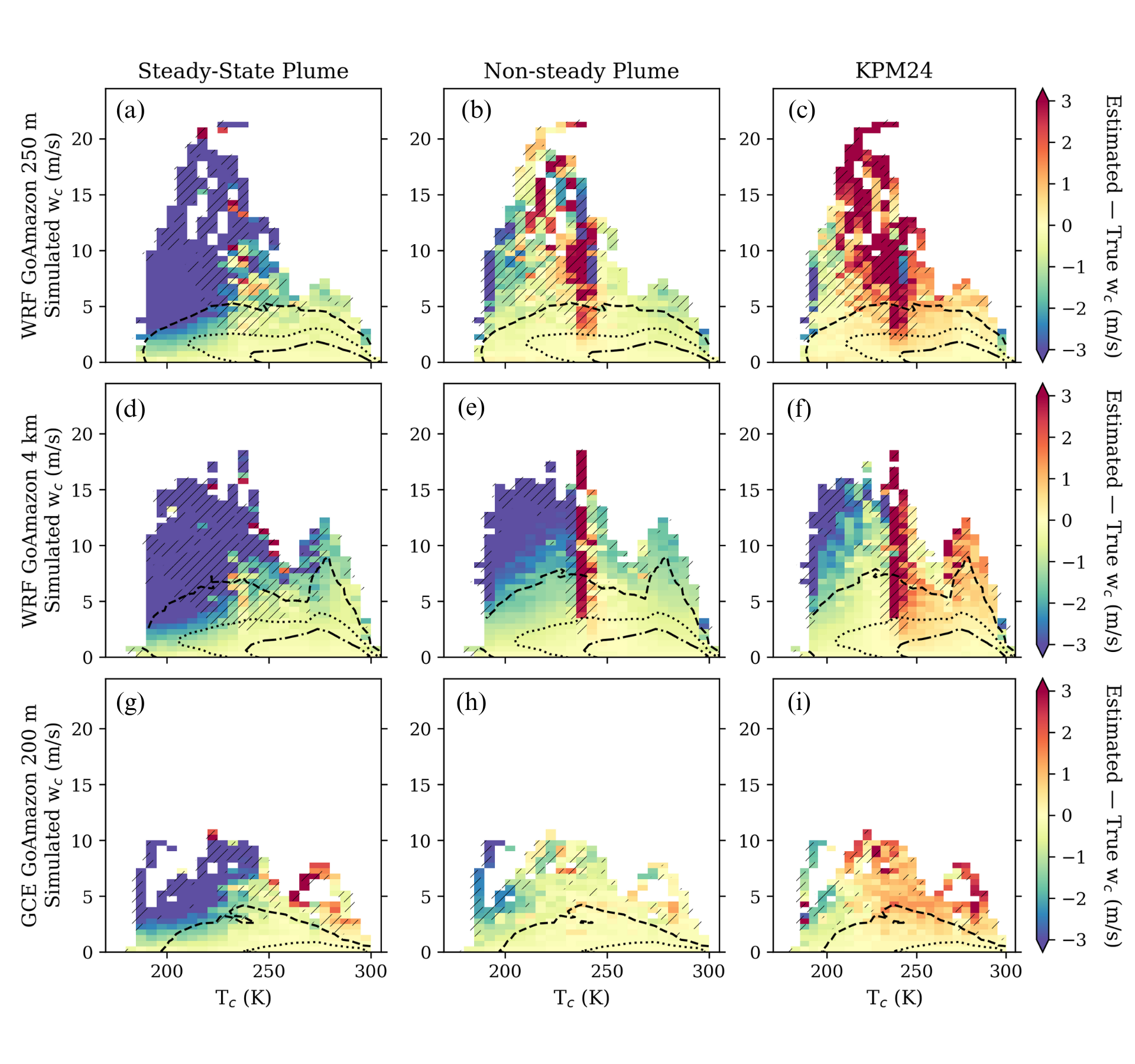}
    \caption{The most likely error in the estimated $w_c$ as a function of $T_c$ and simulated $w_c$. Vertical velocity and temperature bins are 0.5 m/s and 5 K, respectively. The shaded regions mark areas where the interquartile range of the distribution of error is greater than 1 m/s. The dashed, dotted, and dashed-dotted lines mark contours of 100, 1000, and 5000 samples.}
    \label{fig:errors_wc_tc_a}
\end{figure*}

\begin{figure*}[h!]
    \centering
     \includegraphics[width=\linewidth]{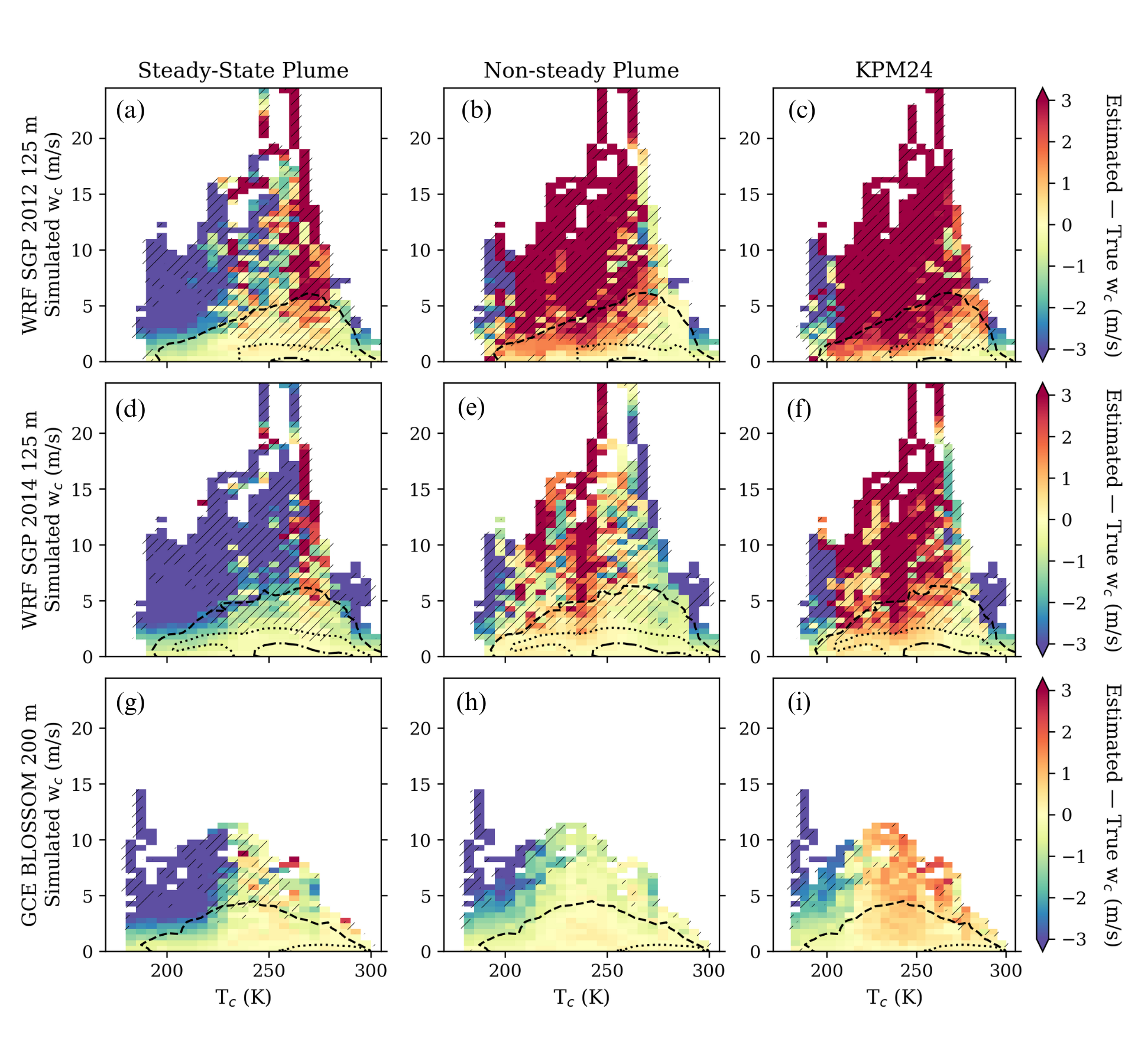}
    \caption{As in Fig. \ref{fig:errors_wc_tc_a}, but for the three mid-latitude simulations.}
    \label{fig:errors_wc_tc_b}
\end{figure*}

We further examine the temperature dependence of $w_c$ estimation errors by plotting the most likely error as a function of simulated $w_c$ and $T_c$ in Figs. \ref{fig:errors_wc_tc_a} and \ref{fig:errors_wc_tc_b}. Broadly, the errors in velocities estimated with $\alpha_p$ are smaller than the ones given by the other two estimators over the widest range of temperatures and simulated vertical velocities. When the simulated velocities are small, however, all of the estimators accurately predict $w_c$. In most cases, the errors increase with increasing simulated vertical velocity. This may be because all of the versions of $\alpha$ assume hydrostatic balance, which breaks down at high velocities. 

At around 240 K, all of the $w_c$ estimates exhibit a transition in the most likely errors, behavior that is most apparent for high simulation velocities. The steady-state model switches from slightly overestimating to underestimating $w_c$ as temperature decreases. In the non-steady-state case, there is a narrow region around 240 K where the most likely error is high and positive, although this is not seen in the GCE results. This may be due to the smaller sample size and the smaller phase space spanned by the simulated $w_c$ values in the GCE models. The errors in $w_c$ estimates given by the KPM24 model peak at around 240 K as well. 

The errors in the $w_c$ estimates given by $\alpha_p^{steady}$ are noisy, but the magnitude of the errors is low when velocities are low. The magnitude of $\alpha_p^{steady}$'s most likely error is less than 1 m/s for a significant portion of samples in both tropical and mid-latitude environments across all of the simulations. Errors are highest when temperatures are below $\approx270$ K and simulated velocities are greater than 2 m/s. There, the velocities estimated with the steady-state plume model are much lower than the simulated ones. 

The errors in the vertical velocities estimated with $\alpha_p$ exhibit higher precision than the steady-state ones. The precision of the vertical velocity estimates decreases in the mid-latitude simulations, and the magnitude of the errors increases. Over most of phase space shown in Fig. \ref{fig:errors_wc_tc_a}, the non-steady plume model estimates $w_c$ to within 1 m/s of the true value. Biases appear at low temperatures and high simulated velocities, where this model tends to underestimate $w_c$. In the mid-latitude cases, there are major differences between the accuracy of $w_c$ estimated in the three different simulations. In the SGP 2012 case, the estimated $w_c$ is much higher than the true one. Similar overestimates are found in the SGP 2014 case, but they are not as pronounced. Across both of the SGP simulations, we find that the spread in the errors in the vertical velocities are large. In the GCE BLOSSOM case, the estimated vertical velocity is fairly accurate. 

The most likely error in $w_c$ estimated using the KPM24 model is small in the tropics, and, as in the $\alpha_p$ case, exhibit higher precision than the steady-state $w_c$ estimates. In the mid-latitude simulations, we again find that the estimated vertical velocities are much higher than the true ones in the WRF simulations. However, velocities estimated in the GCE BLOSSOM case are accurate to within 1--2 m/s almost everywhere. 

The differences in the errors between the tropical and mid-latitude WRF simulations indicate that this method for estimating velocity may be more biased in the mid-latitudes. While the GCE GoAmazon results are comparable to the GCE BLOSSOM results across all of the $w_c$ estimators, the increases in both errors and noisiness in the WRF mid-latitude $w_c$ estimates suggest that the method should be revised to improve reliability outside of the tropics.

\section{Discussion} \label{sec:discussion}

The results of this study suggest that it is possible to obtain fairly accurate estimates of $w_c$ and its uncertainty in the tropics using analytical expressions of $\dot{q}_{vc}/w_c$. The estimates given by $\alpha_p$ and $\alpha_{KPM}$ exhibit especially high accuracy. At low temperatures (or high altitudes), errors increase as the true vertical velocities increase. Overall, this method is promising because the errors in $w_c$ estimates are comparable to typical uncertainties in $w_c$ measurements taken by radar wind profilers, instruments widely used to characterize vertical velocity to within 1--2 m/s \citep{heymsfield_characteristics_2010}. 

The temperature sensitivities of $\alpha_p$ and $\alpha_{KPM}$ illustrate that these expressions will lead to the most precise $w_c$ estimates when $T_c$ exceeds $\approx$ 240 K. When $T_c$ is low, the spread in the velocities expected by shifting the in-cloud temperature by 5 K becomes large, especially when the true vertical velocity exceeds 10--15 m/s. Small uncertainties in $T_c$ will lead to large uncertainties in $w_c$. In the tropics, this increased uncertainty impacts a small fraction of the samples. There, the majority of updraft velocities are less than 10 m/s, as demonstrated in analysis of field campaign and Doppler retrievals \citep{lemone_cumulonimbus_1980, black_vertical_1996, giangrande_convective_2016}, and in the contours delineating the number of samples in Figs. \ref{fig:errors_wc_tc_a} and \ref{fig:errors_wc_tc_b}. However, the most intense updrafts are likely responsible for a significant portion of the vertical mass flux \citep{wang_characteristics_2014} and associated detrainment that forms anvil clouds. 

The anomalous behavior of the $w_c$ estimators between 230 and 240 K is likely related to the numerical implementation of the homogeneous freezing temperature of water. (While it is pronounced in the WRF simulations, this anomaly is absent in the GCE ones, illustrating one way that differences in the simulations may impact the assessment of the $w_c$ estimators. Future work validating this method against more sophisticated microphysics schemes may clarify where these simulation differences come from and further improve the $w_c$ estimates.) In the velocity estimates given by $\alpha_p$ and $\alpha_{KPM}$, the errors within this temperature range are high and positive. This is likely because all liquid turns to ice around 240 K, increasing the latent heating rate. While this should also increase the vertical velocity, the condensate mass may dampen the vertical response of the simulated cloud to the increased heating \citep{varble2023opinion, elsaesser_improved_2017}. The estimated $w_c$ may be too high because the analytical expressions neglected the impact of condensates on vertical motion. 

At high altitudes, many of the assumptions made in deriving $\dot{q}_{vc}/w_c$ are no longer valid and may contribute to the increased error in the plume-based estimated vertical velocities. Since the cloud is primarily ice in this region, deposition, not condensation, is the source of the latent heating. Another assumption that may impact the ability of these models to accurately estimate $w_c$ is the absence of horizontal advection. In \citet{grant_linear_2022}, estimates of $w_c$, which were made by inputting the $\dot{q}_{vc}/w_c$ values extracted from numerical data into a vertical flow equation, improved at high altitudes. This suggests that a more complete model that includes horizontal advection and diffusion may be necessary to improve the vertical velocity prediction at high altitudes. Additionally, radiative heating makes up an increasingly larger fraction of the diabatic heating rate of the cloud as the altitude increases \citep{chen_relationship_2024}, negating our assumption that phase changes constitute the leading order source of heating. Other effects like secondary ice formation, which can impact supersaturation \citep{yano_iceice_2011}, may also be relevant at these altitudes because they may increase the deviation from saturation. 

The influence of entrainment on the estimated vertical velocities was only partially neglected when we set $\epsilon = 0$ and $\bar{\epsilon}=0$ because we numerically calculated the vertical derivative of $T_c$ to find $dT_c/dz$. \citet{peters_generalized_2022} demonstrate that only temperature profiles obtained from $dT_c/dz$ equations that contain entrainment led to accurate buoyancy profiles. They found that errors in temperature accumulated with height when integrating from the surface to the top of the cloud, and were especially large when entrainment was neglected. Our approach differs from that of \citet{peters_generalized_2022} because we apply a thermodynamic budget framework to estimate the instantaneous vertical velocity at each height. In this way, we avoid accumulating errors when integrating in the vertical direction. Moreover, we use the ``observed'' $dT_c/dz$ in our analysis, which includes any impact of entrainment on the temperature lapse rate. 

Despite entrainment assumptions made in the plume model, it may be possible to use $\alpha$ to examine several proposed influences on entrainment rate. These include radial size of the cloud \citep{takahashi_revisiting_2021, peters2023analytic}, environmental humidity \citep{lu_observational_2018}, among other influences \citep{becker_estimating_2018, takahashi_revisiting_2023, yano_well-mixed_2022, warner_microstructure_1969}. The models derived in this work may also provide a means to test different entrainment parameterizations, including stochastic formulations \citep{romps_nature_2010}, although assumptions in the plume model may limit the conclusions.

While this work focused on quantifying vertical velocity, by inverting our procedure it is also possible to estimate latent heating rate given $w_c$. Because latent heating is difficult to measure, the most widely used retrieval algorithms rely on cloud-resolving models \citep{shige2004spectral, tao_trmm_2016}. Vertical profiles of latent heating can be constrained from measured precipitation rates and convective/stratiform fractions \citep{tao_retrieval_2006}. More direct spaceborne retrievals of vertical velocity have begun with the recent launch of EarthCARE \citep{wehr_earthcare_2023}. Combined with the upcoming launches of the INvestigation of Convective UpdraftS (INCUS; \url{https://incus.colostate.edu}, \citet{vandenheever2023}) and Atmosphere Observing System (AOS; \url{https://aos.gsfc.nasa.gov}) missions, a series of satellites that will more directly measure vertical motion \citep{stephens_distributed_2020, dolan_time_2023}, it may be possible to estimate the associated latent heating rate using the expressions derived in this paper.

\section{Summary} \label{sec:summary}

We present and evaluate a method to estimate the mean vertical velocity of convective updrafts from profiles of temperature, pressure, and latent heating rate. The method relies on analytical expressions for the approximately linear relationship between vertical velocity and condensation rate, which we relate to latent heating. Two of the analytical expressions derived in this work are based on a one-dimensional plume model, and the third is derived from the rate of supersaturation in convective clouds, detailed in \citet{kukulies_simulating_2024}. The vertical velocity estimates are validated against tropical and mid-latitude convection simulations using two different simulation types. 

Despite several assumptions made in developing the expressions, such as pseudoadiabtic ascent and the absence of ice, the expressions can be used to estimate the average vertical velocity in convective updrafts with qualitative accuracy. The results suggest that the $w_c$ estimates given by the steady-state plume model are accurate to within a few m/s when the sample size is large, but the spread in the errors is broad. In the tropical simulations, $w_c$ estimates given by the non-steady-state plume model are accurate to within $\approx 1$ m/s in most cases. The KPM24 model exhibits similar accuracy, although it tends to slightly overestimate velocity compared to the non-steady plume model. Both the non-steady plume and KPM24 models overestimate velocity around the homogeneous freezing level of water, where the steady-state plume model is the most accurate. Both tropical and mid-latitude simulations are used in the validation, and the increased errors in the mid-latitude simulations suggest that this method is best suited for the tropics. While both the non-steady-state and KPM24 expressions are sensitive to temperature, potentially leading to a range of $w_c$ estimates if temperature uncertainties are high, these uncertainties are insignificant in the lower troposphere.

Overall, the average $w_c$ profiles in convective cores can be characterized to within a few m/s of the true values, providing a way to estimate updraft vertical velocity using readily available satellite data. Globally quantifying $w_c$ will allow for observation-based estimates of convective clouds' role in the global circulation of the atmosphere and be invaluable for the improvement of weather and climate simulations. 

\acknowledgments
The authors would like to acknowledge funding support from NASA Grants 80NSSC23K0123 and 80NSSC23K0116 awarded to CUNY (ZL), and NASA Grants 80NSSC23K0122,  80NSSC23K0117, and NSF Grant AGS-2149355 awarded to Columbia University-GISS (GE). Additional support was provided from the INCUS project under Grant 80LARC22DA001. This work was also supported by funding from the Columbia Climate School (AD). We thank Dr. John Peters and two anonymous reviewers whose comments helped improve this manuscript.   

%
%
\datastatement

The WRF simulations are available on Globus \url{https://app.globus.org/file-manager?origin_id=87afe1af-7a96-44f3-9adf-9a5d6813d69e\&origin_path=\%2F}. For the GCE simulation data, contact Toshi Matsui at toshihisa.matsui-1@nasa.gov or Amel Derras-Chouk at ad4429@columbia.edu.


%




%



\bibliographystyle{ametsocV6}
\bibliography{references}

\end{document}